\documentclass[conference]{IEEEtran}

\usepackage[pdftex]{graphicx}
\graphicspath{{../pdf/}{../jpeg/}}
\DeclareGraphicsExtensions{.pdf,.jpeg,.png}

\usepackage[cmex10]{amsmath}
\usepackage{array}
\usepackage{mdwmath}
\usepackage{mdwtab}
\usepackage{eqparbox}
\usepackage[caption=false,font=footnotesize]{subfig}
\usepackage{stfloats}
\usepackage{url}
\usepackage{hyperref}

\makeatletter 
\newcommand{\linebreakand}{%
  \end{@IEEEauthorhalign}
  \hfill\mbox{}\par
  \mbox{}\hfill\begin{@IEEEauthorhalign}
}
\makeatother

\begin{document}
\title{Bringing Rust to Safety-Critical Systems in Space}


\author{
  \IEEEauthorblockN{Lukas Seidel\IEEEauthorrefmark{1}}
  \IEEEauthorblockA{Binarly, TU Berlin\\
  lukas@binarly.io}
  \and
  \IEEEauthorblockN{Julian Beier\IEEEauthorrefmark{1}}
  \IEEEauthorblockA{TU Berlin\\
  j.beier@campus.tu-berlin.de}
  \linebreakand
  \IEEEauthorblockN{}
  \IEEEauthorblockA{\IEEEauthorrefmark{1}Both authors contributed equally to this work}
}



%

\IEEEoverridecommandlockouts
\makeatletter\def\@IEEEpubidpullup{2.5\baselineskip}\makeatother
\IEEEpubid{\parbox{\columnwidth}{
    {\fontsize{7.5}{7.5}\selectfont IEEE Security for Space Systems (3S) 2024 \\
    27-28 May 2024, Noordwijk, Netherlands \\
    https://atpi.eventsair.com/24a06---3s2024/}
}
\hspace{\columnsep}\makebox[\columnwidth]{}}

\maketitle
\begin{abstract}
The development of safety-critical aerospace systems is traditionally dominated by the C language.
Its language characteristics make it trivial to accidentally introduce memory safety issues resulting in undefined behavior or security vulnerabilities. 
The Rust language aims to drastically reduce the chance of introducing bugs and consequently produces overall more secure and safer code.
However, due to its relatively short lifespan, industry adaption in safety-critical environments is still lacking. \\
This work provides a set of recommendations for the development of safety-critical space systems in Rust.
Our recommendations are based on insights from our multi-fold contributions towards safer and more secure aerospace systems: 
We provide a comprehensive overview of ongoing efforts to adapt Rust for safety-critical system programming, highlighting its potential to enhance system robustness. 
Next, we introduce a procedure for partially rewriting C-based systems in Rust, offering a pragmatic pathway to improving safety without necessitating a full system overhaul. 
During the execution of our rewriting case study, we identify and fix three previously undiscovered vulnerabilities in a popular open-source satellite communication protocol.
Finally, we introduce a new Rust compiler target configuration for bare metal PowerPC.
With this, we aim to broaden Rust's applicability in space-oriented projects, as the architecture is commonly encountered in the domain, e.g., in the James Webb Space Telescope. 
\end{abstract}

\section{Introduction}

Space exploration and utilization are undergoing a transformation, characterized by an increasing diversity of spacecraft, ranging from large, sophisticated systems such as space telescopes to smaller, more accessible and flexible platforms such as CubeSats. 
These compact satellites, particularly prevalent in Low Earth Orbit (LEO), exemplify the trend towards miniaturization and democratization in space technology. 
Significantly higher affordability and accessibility as well as short development life cycles led to their number doubling between 2019 and 2022~\cite{UNOOSA2022}.
A lot of spacecraft rely on software predominantly written in C, a language known for its performance and mature ecosystem but not for its safety features. 
This reliance raises critical concerns, especially as these systems play increasingly vital roles in scientific research, communication, and Earth observation.

While C's dominance in aerospace systems is rooted in its historical presence in embedded systems, its inherent lack of memory safety and the manual management of resources has led to a plethora of vulnerabilities and safety issues. 
Such issues can directly impact the reliability and security of space missions. 
Although long called for, the security of space systems was neglected for a long time~\cite{vacuum}, resulting in unsafe systems and even creating the opportunity for unauthorized takeover~\cite{odyssey}.
As the sector evolves, with a growing emphasis on smaller, cost-effective satellites and more frequent launches, the need for robust, secure, and reliable software becomes increasingly important.

Rust, a modern programming language that prioritizes safety and performance, provides a possible alternative. 
Rust's design inherently eradicates a whole class of common bugs found in C programs, particularly those related to memory safety. 
By leveraging Rust's capabilities, it is possible to significantly reduce the incidence of system failures and security vulnerabilities, ensuring higher standards of reliability for these critical systems.
Introducing Rust to large-scale software projects such as Android showed that less memory-unsafe code correlates closely to fewer memory safety vulnerabilities, dropping from 76\% to 35\% of Android's total vulnerabilities between 2019 and 2022~\cite{RustAndroid}.
Although its language features were theoretically evaluated for their applicability in safety-critical environments in the past~\cite{UKMOD}, Rust is yet to be widely adopted by the industry.
Practitioners often prefer standardized languages while Rust did not have a language specification for the longest time, making it harder to talk about problems such as undefined behavior.
Additionally, especially safety-critical parts of industrial space-grade systems might require qualified software toolchains.
The shift to a memory-safe language like Rust aligns with the increasing complexity and diversity of embedded systems in space missions.
At the same time, replacing all legacy C code is not feasible.
Instead, more and more memory-safe language implementations should gradually be used or notoriously error-prone code sites should be partially replaced.

In this work, we make several contributions to improve the safety and security of critical onboard systems in space by leveraging Rust.
First, we review current developments toward enabling Rust in safety-critical contexts.
We put our focus on practical work and implementations, providing recommendations to practitioners on how they can utilize Rust in an environment with high safety standards.
We discuss the state of the embedded Rust ecosystem, as we focus on system programming, e.g., for satellite firmware, efforts toward Rust standardization and qualification as well as ways to integrate Rust code in existing C projects.
To this end, we conduct a case study on how to replace existing C code on the function level with new Rust implementations without changing the user interface or introducing unnecessary dependencies.
The result of the introduced process is a library that can transparently replace the original one.
The end user, who is developing on top of the original library, does not need to make any changes to their code.
Finally, we showcase a new target for the Rust compiler, bare metal PowerPC.
Bringing embedded Rust to this target contributes to Rust's applicability in space systems, as PowerPC is still a common architecture in radiant-resistant platforms.

In summary, we make the following contributions:
\begin{enumerate}
    \item We evaluate the state of the Rust ecosystem for use in safety-critical systems in space.
    \item We present a process to partially replace components of software developed in C with equivalent Rust implementations. 
    \item We identify and patch three security issues in the \textit{Cubesat Space Protocol}, a popular open-source packet communication protocol for satellites. 
    \item We develop a new target configuration for the Rust compiler, allowing the compilation of programs for bare metal PowerPC CPUs. 
    \item Based on the combined insights of our contributions, we develop a set of recommendations for practitioners in the realms of space system development.
\end{enumerate}

\section{Background}

\subsection{The Rust Programming Language}
Rust was first introduced in 2006 by the Mozilla Foundation with its version 1.0 release announced in 2015.
The language has since been adopted rapidly, with its strong focus on code safety and high performance comparable to that of C or C++ as the main reasons for its success~\cite{NatureRust, SOSurvey}.

Rust features a strong type system and enforces memory safety guarantees, adding to the language's safety~\cite{RustBook}.
Guarantees include that there is only one mutable (writeable) reference to an object or several readable ones, but not both at the same time.
This ownership system introduces zero runtime overhead as it is enforced at compile time and effectively eliminates a large class of correctness errors many C implementations are suffering from.
Classic memory safety bugs that are typically avoided with Rust include buffer overflows, use-after-frees and null pointer dereferences~\cite{RustAndroid}.

Using the \texttt{unsafe} keyword, it is still possible to perform potentially unsafe operations where necessary~\cite{RustBook, UnsafeRust}.
Raw pointer accesses are one example of an unsafe operation as the borrow checker is not able to reason about these without a type-safe view on the underlying memory.
Especially in embedded systems and low-level programming, such operations often are not completely avoidable.
Usage of \texttt{unsafe} blocks can often be restricted to very few code sites, e.g., by wrapping unsafe functionality in safe interfaces.
Overall, Rust offers full control where needed while still being considerably safer than the alternatives, making it highly suitable for systems programming~\cite{SafeSystemsRust}.
Besides C and assembly, it is the only language supported for Linux kernel development.

\subsection{Security Issues in Space}
Security and safety concerns in software engineering are closely related~\cite{cybersafety}.
Security issues such as memory corruptions can result in unwanted changes to internal state representations and can have consequences such as hardware damage or loss of control.

Both from an information security- but also a regulatory standpoint, cybersecurity in space has been increasingly discussed recently~\cite{odyssey,Jacobs2023} after not being a major concern for a long time~\cite{vacuum}.
Although attacks on in-orbit systems such as satellites have a long history~\cite{steinberger2008}, many communication systems still have no access control measures in place~\cite{odyssey}.
They rely on the flawed principle of security-by-obscurity, as operators put their trust in the high barrier to entry to even establish communication with a system in space.
Inexpensive off-the-shelf components and open satellite and antenna designs have drastically lowered this barrier in recent years.
This increasingly allows attackers to access in-orbit systems, further broadening the attack surface when they can send (malformed) packets that get parsed or even directly issue telecommands.
A recent survey on satellite security found that, while hardware-software system integration testing and unit-testing are quite common in ensuring the correctness of space-grade systems, most satellite development does not include dedicated security testing~\cite{odyssey}.
The authors of the survey uncovered memory corruption-based vulnerabilities in critical components such as the Command and Data Handling System (CDHS) firmware.
Memory corruptions will often lead to crashing or otherwise ill-behaving systems and hence are large safety issues.
At the same time, common system-level security mitigations such as non-executable stacks or Address Space Layout Randomization (ASLR) are mostly absent in embedded systems and Real-Time Operating Systems.
The lack of basic security mitigations turns many bugs into exploitable vulnerabilities.

More recently, fuzzing has been increasingly applied to space systems.
Fuzzing is an automated software testing approach in which a system under test is repeatedly presented with automatically generated inputs~\cite{libafl}.
Presenting a program that expects structured input with unexpected formats or random data aims to find unhandled edge cases resulting in bugs.
Analyzing such findings and fixing their root causes can help to reduce bugs and lead to improved software quality.
The technique has been successfully applied to find security vulnerabilities in a multitude of different systems~\cite{nyx, metaemu} and increasing its applicability and efficacy are ongoing research efforts~\cite{fuzz4all, 2023_safirefuzz}.
As part of a security analysis of COSPAS-SARSAT, a satellite-aided search-and-rescue initiative, Costin et al. fuzzed EPIRB protocol implementations~\cite{DBLP:conf/spacesec/CostinTK023}. 
Scharnowski et al. perform a case study of fuzzing embedded space systems and identify bugs in the firmware of three satellites~\cite{DBLP:conf/spacesec/ScharnowskiBWH23}.
As systems in space are inherently physically inaccessible, resetting the systems often is non-trivial.
In the worst case, security issues like Remote Code Execution can result in an unauthorized takeover and permanent loss of control over a space vehicle~\cite{odyssey}.

\subsection{Safety-Critical Systems}
Designing safety-critical systems is especially important in sectors such as aviation, automotive and industrial control systems, where failure of a system will result in serious damage to expensive equipment or even persons.
Safety-critical system requirements aim to improve reliability.
Typical measures for this are added redundancy in a fault-tolerant system or fail-passive designs.
For the scope of this work, we also treat mission-critical and security-critical considerations as parts of "safety-critical".
While faulty software, e.g., in unmanned space flight, might not directly endanger human lives, it can still lead to unintended behavior, resulting in failure to achieve certain mission goals, damaging expensive hardware, loss of sensitive proprietary data or even loss of access to remotely controlled systems.
Developing not only safe but also secure systems is important to avoid such unforeseen consequences.

Different standardizations exist to assess the safety requirements of a system in a given context and to consequently provide guidelines for safety-aware and risk-minimizing implementation.
In the following, we will give a concise overview of relevant industrial certifications for safety-critical systems.
The most common standard for software in aerospace systems is the RTCA DO-178C, the \textit{Software Considerations in Airborne Systems and Equipment Certification}~\cite{RTCA2011DO178C}.
It provides guidelines for the development of aviation software, ensuring safety and reliability in compliance with regulatory requirements.
The ISO 26262 standard for "Road vehicles - Functional safety" defines multiple \textit{Automotive Safety Integrity Levels} (ASIL) for electronic systems in serial production vehicles~\cite{ISO26262}.
An initial risk assessment establishes the ASIL for a given product, from A to D with D being the highest.
The levels define the safety requirements necessary to be qualified under ISO 26262.
Having increasingly high integrity requirements, they aim to avoid unreasonable residual risks in functional safety.
The IEC 61508 standard covers functional safety for safety-related electronic systems in industrial environments~\cite{IEC61508}.
Each safety function is assigned a target Safety Integrity Level (SIL) with 4 being the highest. 
The SIL is quantified by the minimum safe-failure fraction and the maximum probability of dangerous failure.
For every new environment, an individual definition of what \textit{dangerous} means is required, usually in the form of requirement constraints. 
Ultimately, the standard includes methods on how to design and maintain automatic safety-related systems.
It is applicable in all industries.

All of these standards usually deal with safety \textit{lifecycles}, including development, production and operation.
Usually, safety-critical software engineering is comprised of standardization-guided coding and system analysis, manual inspections, documentation, software testing and verification~\cite{8006260}.
These steps can include the usage of methods and components specifically certified for certain domains, e.g., formally verified compilers.

\section{Developing Safe Space Systems in Rust}
The majority of software needed to operate a spacecraft relies on the C programming language, especially in (embedded) systems development.
Consequently, software engineers' expertise in the field historically is mostly focused on this legacy ecosystem and C as a language.
This leads to further reluctance to adopt a new and relatively young programming language in a very change-averse space.
With the additional high degree of requirements to meet in the highly regulated area of space systems~\cite{NASA2003}, industry adoption of Rust is still in its infancy.
In open-source space software, Rust is already being used: In KubOS~\cite{kubos}, a software stack for satellites, component functionality is largely implemented in Rust.
FreeRTOS, a real-time Operating System (RTOS) prominently used in in-orbit systems~\cite{FreeRTOSinSpace}, offers mature Rust bindings. 
Although such efforts demonstrate the space community's willingness to adopt a safer alternative to the ubiquitous C, real-world (industry) usage is still limited due to safety concerns and the need for legacy interoperability.
At the same time, Rust's focus on safety has been shown to lead to fewer security- and safety-relevant bugs in large-scale real-world systems~\cite{RustAndroid}.
As Rust not only offers language features enabling safe systems programming~\cite{SafeSystemsRust} but also high interoperability with existing C code, we see Rust as a promising candidate to replace C in the future engineering of safety-critical embedded systems.

\subsection{The Embedded Rust Ecosystem}
\begin{figure}[t!]
    \centering
    \includegraphics[width=\linewidth]{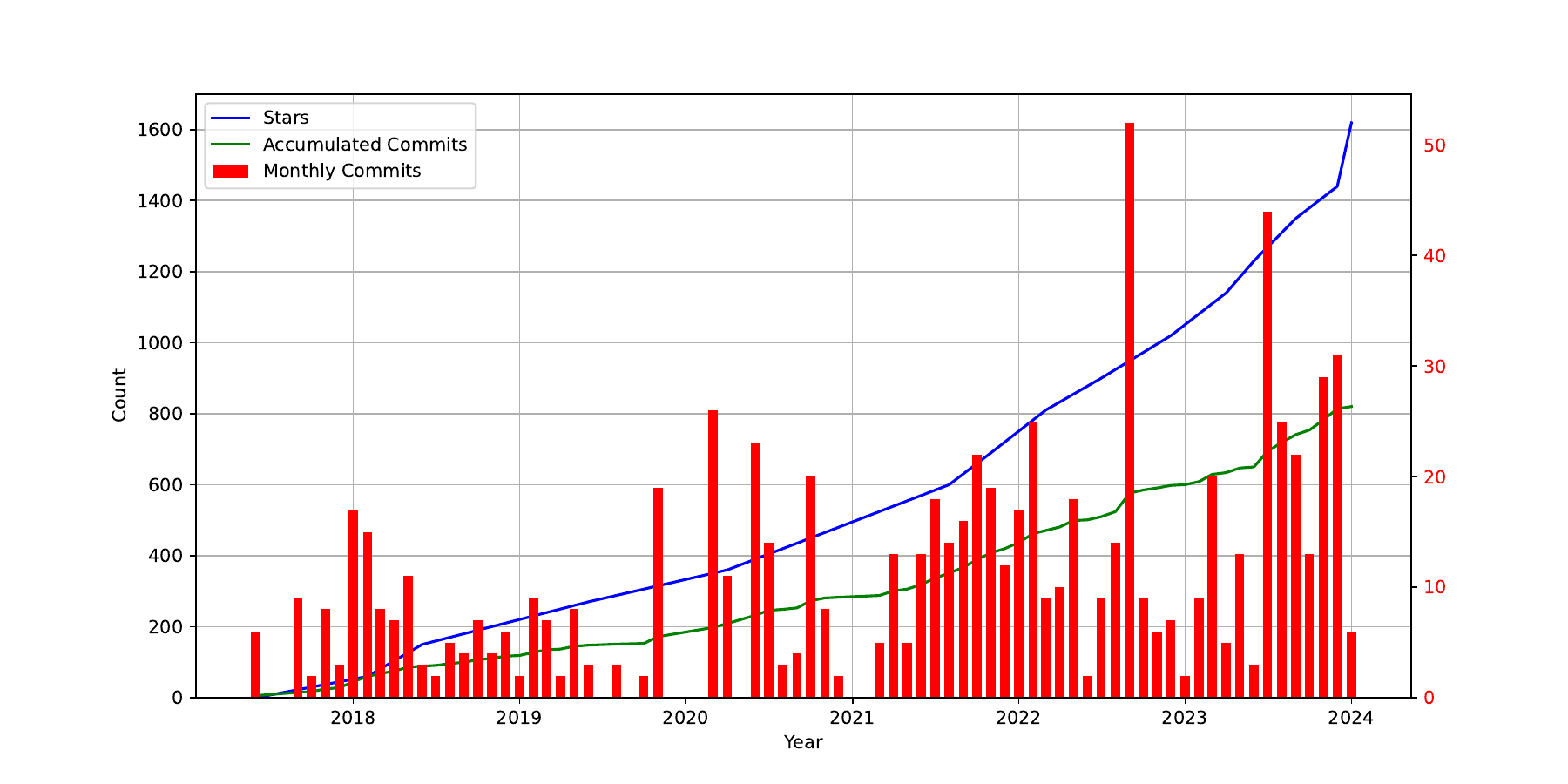}
    \caption{Popularity of the \texttt{embedded-hal} Rust crate. The GitHub repository shows steady growth in popularity, measured in GitHub Stars, while the amount of monthly contributions also increases year-over-year.}
    \label{fig:embedded-hal-act}
\end{figure}

A programming language's ability to manage system resources and interface with hardware and real-time systems is an important aspect when considering its applicability in safety-critical system programming.
Spacecraft such as satellites usually integrate multiple embedded systems for their various tasks~\cite{odyssey}.
One important aspect of embedded system development is the support of the diverse hardware components to be found on microcontrollers.
Whether a programming language is suitable for a given task also might depend on whether there are drivers or other hardware interfaces available to achieve a given task.
Commonly, hardware abstractions are used to write device-independent drivers for embedded systems.
A driver can be any piece of code interacting with external peripherals, such as sensors, antennas or actuators.
By developing the driver on top of a \textit{Hardware Abstraction Layer} (HAL), it is agnostic of the underlying platform and can be used without modifications on a range of embedded devices, e.g., AVR or ARM Cortex-M microcontrollers.
For Rust, the popular \texttt{embedded-hal} package (crate) offers this foundation and acts as an integral part of the embedded Rust ecosystem~\cite{embeddedhal}.
There are currently nearly 200 drivers built on top of this HAL, ranging from support for humidity sensors over gyroscopes and radio transceivers to embedded-graphics-compatible LCD drivers~\cite{drivers}.
By building upon a common HAL, these drivers currently support 48 different microcontrollers that have \texttt{embedded-hal} implementations.
There are also whole firmware projects realized with \texttt{embedded-hal} as a base:
\textit{Stabilizer} is a firmware for a Digital Signal Processor for quantum physics experimentation, including telemetry via MQTT~\cite{stabilizer}.
The realization of such complex projects in Rust is a strong signal for embedded Rust's efficacy and maturity.
Development of this unified Rust HAL started in 2017 and contributions and popularity are steadily growing (cf. Figure~\ref{fig:embedded-hal-act}), showing the embedded community's interest in Rust.
The crate was recently released in version 1.0\footnote{https://blog.rust-embedded.org/embedded-hal-v1/}, stabilizing all traits and committing to not introducing breaking changes in the future.
Other projects of the \textit{Rust on Embedded Devices Working Group} include the \texttt{heapless} crate~\cite{heapless}.
The crate offers implementations of common data structures such as vectors or double-ended queues not depending on a dynamic allocator, thus being especially suited for embedded development.

Systems with ARM Cortex-M CPUs, such as STM32 microcontrollers, are a common component of space systems such as CubeSats~\cite{DBLP:conf/spacesec/ScharnowskiBWH23}.
The Rust ecosystem for Cortex-M-based systems is especially mature.
A collection of Cortex-M crates offers support for peripheral access, interrupt handling and startup sequences~\cite{cortexm-crates}.
They provide safe and idiomatic Rust interfaces to potentially unsafe functionality where possible.
There are also recent developments explicitly aimed at further enhancing the security and safety of embedded systems.
The \textit{flip-link} tool for Cortex-M systems addresses the issue that in certain bare metal environments even code without any \texttt{unsafe} blocks may not be memory-safe if a stack overflow, i.e., undefined behavior, is present~\cite{flip-link}.
A downwards-growing stack might collide with the \texttt{bss} / \texttt{data} memory region, overwriting static variables and resulting in undefined behavior.
The tool flips the memory layout and positions the stack below the static memory regions.
If the stack now overflows, a catchable hardware exception will be thrown when attempting to read beyond the physical RAM boundaries.

\subsection{Safety-Critical Software Engineering}
In 2022, researchers of the \textit{United Kingdom Defence Science and Technology Laboratory} proposed evaluation criteria based on RTCA DO-178C to review how well-suited a programming language is for usage in safety-critical contexts in the air domain~\cite{UKMOD}.
For Rust, they concluded that, while some aspects of the language require special attention, there are no major barriers to its use in safety-critical
software in aviation.

\paragraph{Programming Guidelines}
While fully formally verifying a whole ecosystem is infeasible, for the usage in safety-critical contexts certain practices can be mandated.
Standards such as DO-178C do not provide a coding standard in themselves but require that one must be used~\cite{RTCA2011DO178C}.
One of the most popular and up-to-date standards for this use case is MISRA-C~\cite{MISRA2012}.
MISRA-C provides guidelines for the development of safe, secure and reliable embedded systems.
The guideline also acknowledges that software is neither safe nor reliable if it exposes security vulnerabilities, even if the system is functionally correct otherwise.
Originally intended for use in the automotive industry, the standard is nowadays widely used in safety-critical systems and is also the foundation of the NASA Jet Propulsion Laboratory C Coding Standards.
In 2022, the \textit{High Assurance Rust} initiative introduced a programming guide loosely based on MISRA-C~\cite{high_assurance_rust}.
The \textit{High Assurance Rust Book} provides an introduction to systems programming and low-level software security with an emphasis on code safety and functionality.
It provides a taxonomy of MISRA rules, showing that a significant share of rules is automatically enforced by the Rust compiler.
Topics of the book include data structures for resource-constrained embedded environments and static program verification.
The practical examples introduce a safe subset of Rust based on MISRA-C directives.
Furthermore, the book discusses techniques such as differential fuzzing and deductive verification to verify properties Rust's compiler cannot automatically prove.

\paragraph{Standardization and Qualification}
The reliability and determinism of a language's compiler are paramount when developing software for system-critical systems.
Ferrous Systems recently qualified their open-source Rust compiler \textit{ferrocene} under ISO 26262 (ASIL D) and IEC 61508 (SIL 4)~\cite{ferrocene}.
This effort presents the first qualification of Rust for usage in safety-critical systems.
Currently aimed at the automotive and industrial sectors, a future qualification for aerospace systems under DO-178 is planned.
The qualification of the compiler effectively means that it was tested to produce safe results, i.e., given source code implementing a functionality, the target binary is achieving that functionality.
To this end, a diverse set of configurations was tested, e.g., the compiler in a certain version, on ARMv7-M, in a FreeRTOS setting with specific compiler flags.
Libraries such as \texttt{core} require further qualification to assert that they implement their specification.
Although this presents an apparent gap in safety qualification, it is interesting to note that mature languages such as C and C++, although having qualified language specifications, also do not have qualified standard libraries. 
Ferrous System's qualification efforts also produced the first full language specification for Rust~\cite{ferrocene-spec}.

AdaCore recently introduced GNAT (GNU NYU Ada Translator) Pro for Rust~\cite{gnatrust}.
GNAT is a compiler for the Ada programming language based on GCC infrastructure.
Ada is extensively being used in real-time and embedded systems in, e.g., aerospace, and defense industries, aiming for enhanced safety and reliability.
The support for Rust allows the integration of Rust code into existing C, Ada and SPARK projects, introducing bi-directional Ada-Rust bindings.
This integration into the AdaCore ecosystem would also allow for the use of its other features, such as formal verification methods with SPARK, e.g., for communication protocols~\cite{recordflux}.
GNAT Pro for Rust provides versions of \textit{rustc} (compiler), \textit{cargo} (package manager and \textit{rustc} wrapper) and \textit{gdb} (debugger) with long-term support.
Yearly updates include testing and qualification of changes coming from upstream Rust, providing a more stable development environment.
AdaCore also stated that they intend to get language library subsets of \texttt{core} certified, e.g., under DO-178.  

\section{Making Software for Space Systems Safer}
In the following, we will investigate how to increase the overall safety of existing systems by rewriting especially safety-critical parts in Rust.
To this end, we conduct a case study on the Cubesat Space Protocol (CSP), a network protocol for standardized communication in a distributed embedded system such as in CubeSats~\cite{libCSPdocs}.
With over 40 contributors, 2000 commits and 400 stars, it is the most popular open-source library for satellites on GitHub.
Features of the protocol include a thread-safe socket API, a modular network interface system, a small memory footprint and integrations for Linux as well as for popular embedded OSs FreeRTOS and Zephyr.
CSP is being used in orbit: the SUCHAI nanosatellite flight software~\cite{SUCHAI} uses libCSP for communication between subsystems and with the ground station~\cite{suchaiGH}, the European Space Agency (ESA) uses FreeRTOS and libCSP in their 2019 OPS-SAT CubeSat and the German Space Operations Center used the CSP for its \textit{CubeL} CubeSat in 2020~\cite{gsoc}.

\subsection{Security Analysis}
To gain a deeper understanding of the protocol, we conduct a partial security analysis of its C implementation~\cite{libCSPGH}.
Investigating the commit history and past bug fixes of the project showed that the interface implementations inhabit the highest count of high-impact vulnerabilities such as buffer overflows.
This can be explained by the library's zero-copy buffer- and queueing system.
Once a raw packet arrives at an interface and is placed in a queue for further processing, no more raw memory operations are performed on it.
This reduces the overall attack surface on higher layers significantly.
Consequently, we deem the interface backend implementations especially interesting for further analysis and for re-implementation in Rust.

We set up fuzzing for libCSP with \textit{LibAFL}~\cite{libafl}, a highly customizable fuzzing library implemented in Rust.
To fuzz the CAN interface backend, we write a custom harness that allows us to put fuzzer-generated pseudo packets into the queue and test if certain parts of the library crash.
We identify three previously undiscovered security issues:
First, an off-by-one error allows processed CAN frames to trigger a buffer overflow despite the presence of a guard condition.
Second, an unchecked size field in raw incoming CAN frames enables maliciously crafted frames to trigger another buffer overflow.
Third, the missing check of a return value of a \texttt{malloc} call in the USART interface, potentially leading to a NULL pointer dereference.
All findings were disclosed to and acknowledged by the maintainers. 
Jointly, we developed patches and improved the overall code safety of the library
\footnote{https://github.com/libcsp/libcsp/pull/510}.

Partially rewriting a low-level component such as the CAN interface in Rust does not magically make all security considerations obsolete.
The low-level interfaces, e.g., for the CAN bus or ethernet, make use of drivers to retrieve raw data.
This raw data subsequently is assembled into CSP packets and placed in a queue, actions that require raw memory access.
Thus, the usage of \texttt{unsafe} in Rust would be hardly avoidable.
But we agree that, as has been argued before~\cite{RustAndroid}, the use of \texttt{unsafe} blocks can lead to increased caution while implementing such code sections, potentially leading to fewer programming mistakes.
At the same time, multiple bug classes we encountered during the analysis of the library's history would have been trivially avoided in safe Rust.
These include double frees and potential NULL pointer dereferences due to missing checks on the return values of manual memory allocations.

\subsection{Partially Rewriting C Applications}
Interoperability between Rust and C is in a very mature state.
Tooling such as \textit{cbindgen}~\cite{cbindgen} and \textit{bindgen}~\cite{rust-bindgen} can automatically generate bindings from Rust to C and vice-versa.
Using functionality from existing C libraries in new Rust projects or integrating Rust libraries in C projects are common use cases and guides explaining the process can be found plenty.
Many popular libraries such as \texttt{openssl}~\cite{opensslRSGH} use \textit{bindgen} to expose C functionality to Rust programs.
Yet, resources describing the replacement of existing C code with a Rust implementation and linking both back together, involving bidirectional library dependencies, are missing.
We aim to fill this gap and conduct a case study on libCSP.
Reimplementing entire C code bases in Rust quickly becomes infeasibly time-consuming with project size, so we aim to only rewrite chosen components.
Gradually reimplementing parts of a project in Rust is often more realistic and can hence net short-term security benefits.
More critical subsystems can be rewritten before less important ones, with the entire library still functioning as before.

\begin{figure}[t!]
    \centering
    \includegraphics[width=\linewidth]{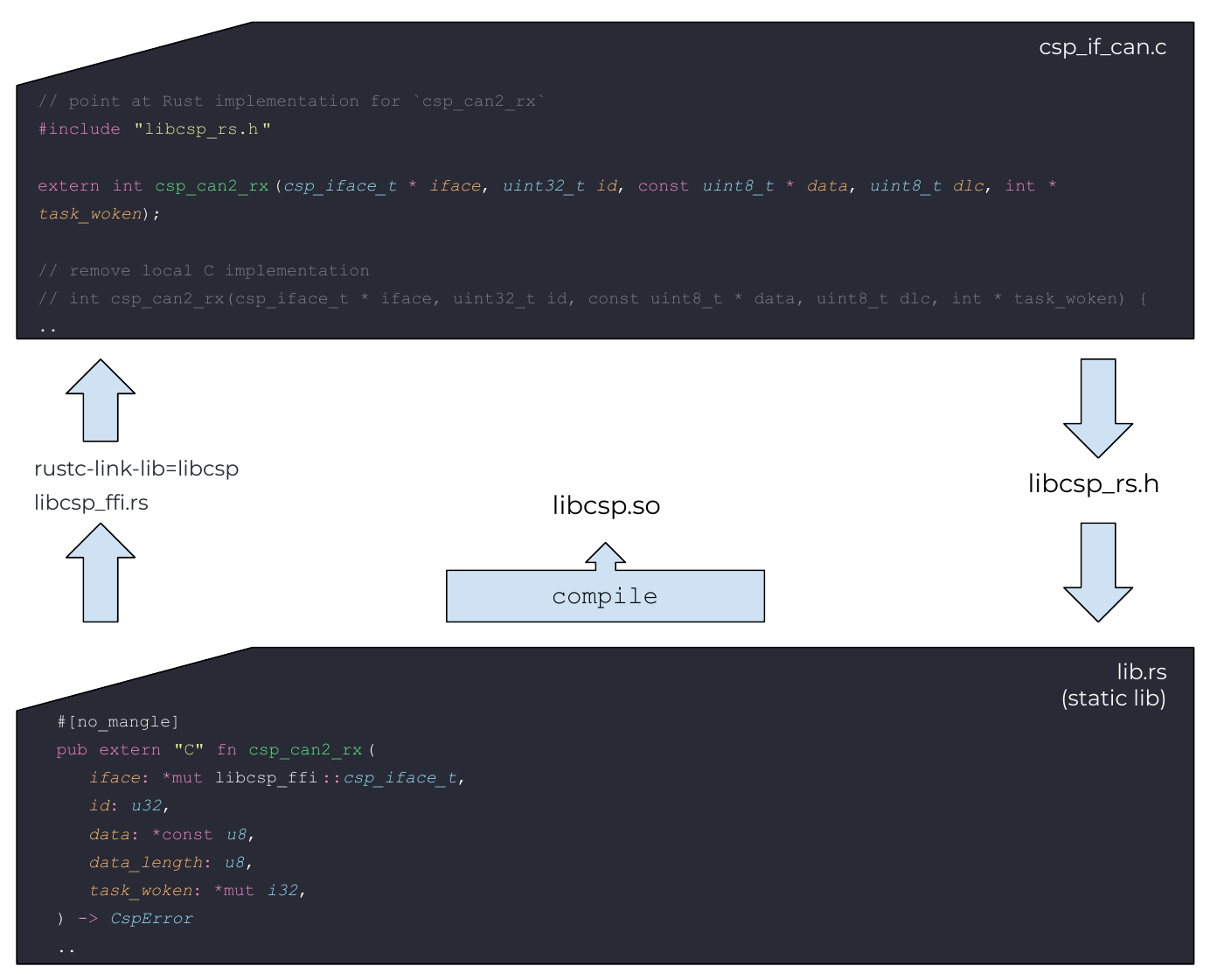}
    \caption{Partial Rewriting and Linking Process. We substitute a single function of libCSP's C implementation with a Rust version and compile it into a single library without the changes being visible to end-users..}
    \label{fig:partial-rewriting}
\end{figure}

To this end, we perform the following steps:
As a target, we choose the \texttt{csp\_can2\_rx} function of libCSP's CAN interface.
We first remove the code in question from the C source code of libCSP, then we mark the function as \textit{extern} to tell the compiler that the function will be available, just not in the C source.
We implement the functionality as a Rust library.
As the function makes calls to other libCSP functions, we use \textit{bindgen} to generate Foreign Function Interface (FFI) bindings.
These bindings include public constants such as \texttt{CSP\_BUFFER\_SIZE}, struct definitions and external C function bindings.
For the most important structs and error types, we add manual definitions and enums, overall aiming to write more idiomatic Rust.
Because the Rust compiler cannot assure the usual safety guarantees across the FFI boundary, every call to a C function from Rust must be marked \texttt{unsafe}.
We build the library as \texttt{no\_std} and \texttt{panic = "abort"} to eliminate outside dependencies and make the integration as minimal as possible.
Our \texttt{build.rs} file contains linker arguments to specify the location of the original C library.
We first compile our static Rust library, linking against the original C headers.
Subsequently, we write a simple C header file defining the interface of the replaced function, which is of course identical to the original one but now an external function.
The process is illustrated in Fig.~\ref{fig:partial-rewriting}.
Finally, we compile libCSP with \texttt{csp\_can2\_rx} marked as extern and linking against our Rust library with the new header file.
The result is a \texttt{.so} file that includes all functionality from the original libCSP with some functions being written in Rust.
To conclude our case study, we compile a test application from the libCSP repository with our library.
Running it confirms that the original functionality is maintained.

One limitation of only partially replacing an implementation with Rust is the need to maintain interoperability. 
This usually means passing raw pointers between FFI functions and using primitive types from Rust's \textit{libc} crate.
While this does not impose limitations on the applicability per se, it potentially limits the use of more idiomatic, and thus, safer, Rust.
Performance-wise, this does not introduce any inherent overhead as FFI calls in C and Rust are just normal function calls after linking.
At the same time, translating data structures between rather primitive C representations and more verbose Rust ones with complex types could lead to a slight performance impact while allowing for the usage of more idiomatic and expressive Rust patterns. 
Overall, our case study demonstrates the feasibility of the approach and we are confident that the introduced process can contribute to faster adoption of Rust in existing code bases, ultimately leading to safer software.

\section{A New Target Platform for Rust}
As our final case study, we showcase how to compile Rust code for a previously unsupported bare metal platform.
This enables embedded Rust development for additional platforms commonly encountered in space systems.

\subsection{Cross-Compilation in Rust}
The \textit{rustc} compiler offers accessible cross-compilation~\cite{RustBook}.
Various target platforms can be specified via a \textit{target triple} consisting of the architecture,
the vendor, the OS type and optionally an environment.
For instance, \texttt{thumbv7m-none-eabi} specifies the bare metal ARMv7-M architecture as the target, as found on ARM Cortex-Ms. 
For code generation, \textit{rustc} relies on LLVM.
Rust specifies three tiers of platform support for compile targets.
Tier 1 is full support, coming with official binaries and automated testing after all changes introduced to the language.
Example platforms for tier 1 support are 64-bit macOS or ARM64 Linux.
Tier 3 targets can be built with Rust but do not come with any official builds or testing from the Rust project.

\subsection{A Bare Metal 32-bit PowerPC Target}
The \textit{Power Instruction Set Architecture (ISA)} features a reduced instruction set computer (RISC) instruction set.
The architecture is of special importance in the space domain as multiple popular radiation-resistant onboard systems and cores are developed on top of it.
The RAD750 single-board computer was released in 2001~\cite{Burchin2002Rad750} and is used in the 2011 Mars Curiosity Rover and the 2021 James Webb Space Telescope.
It is based on the PowerPC 750 core and operates in environments with up to 100,000 rads.
With \textit{P2020 Space}, Teledyne develops a modern radiation-tolerant processor based on PowerPC e500v2 CPUs~\cite{p2020}.
While Rust supports multiple Linux and BSD flavors for PowerPC, there is no support for \textit{bare metal} PowerPC yet.
To fill this gap, we develop a \textit{rustc} target configuration~\cite{customtarget}.

The configuration specifies options for the Rust compiler to be aware of.
These settings include the LLVM-internal target triple, the size of the biggest atomic type of the platform, the target's endianness and pointer width.
For compilation and linking, we rely on pre-built \textit{musl} toolchains from bootlin
\footnote{https://toolchains.bootlin.com/releases\_powerpc-e500mc.html}
and specify the linker in \texttt{.cargo/config}.
After finalizing the target configuration file, we can use \textit{cargo} to compile our program for bare metal PowerPC:
\texttt{cargo build --target=ppc32-unknown-none.json} \\
To evaluate our target configuration, we develop a simple testing application.
The bare metal binary is configured with \texttt{no\_std} and \texttt{no\_main}.
After starting, it uses Rust's \texttt{core::fmt::Write} and C's \texttt{extern "C" fn putchar(ch: i32)} to print out a simple \textit{"Hello World from Rust!"} in a loop.
Finally, we execute the resulting bare metal binary in QEMU, emulating an e500mc CPU: \texttt{qemu-system-ppc64 -nographic -M ppce500 -cpu e500mc ./ppc\_test}
\section{Recommendations}
We condense the insights from our contributions into a set of recommendations for practitioners designing and developing safety-critical embedded space systems.

\textbf{R01. Gradually incorporate Rust:}
Using a memory-safe language such as Rust for (embedded) systems programming can prevent multiple bug classes.
While rewriting a whole code base in a new language in most cases is an inappropriately large effort, writing new components in one should be the go-to way.
This approach is further facilitated by the mature options to interface between Rust and C components.
For especially critical components, it should be evaluated whether a partial Rust rewrite could be beneficial.
Programming guidelines such as \textit{High Assurance Rust}, although incomplete as of writing, and \textit{The Embedded Rust Book} can further help in transitioning a development team and project to Rust.

\textbf{R02. Use qualified toolchains:}
A compiler qualified for safety-critical domains facilitates the acceptance of Rust as a C replacement in projects required to adhere to certain standards.
Qualification of standard libraries such as \texttt{core} will be even more impactful, paving the way to fully qualified Rust programs.
The commercial solutions we mentioned also provide a more stable foundation in the context of relatively fast-changing Rust versions and features and even offer support packages.

\textbf{R03. Utilize the existing embedded Rust ecosystem:}
The embedded Rust community is very active and produces high-quality components such as the \texttt{embedded-hal} and \texttt{heapless} crates.
Using existing solutions where possible accelerates development by granting access to a diverse set of device drivers without the necessity to implement all low-level interfaces and protocols again.
Furthermore, existing implementations can act as examples for Rust-idiomatic system programming. 

\textbf{R04. Employ dedicated security testing:}
Security and safety considerations go hand in hand.
Security-focused testing can uncover issues beyond the scope of traditional unit- and integration testing.
Dynamic approaches such as fuzzing and formal verification through symbolic execution can help to verify properties that static analysis and the Rust compiler cannot automatically assert.
\section{Related Work}
Two ongoing ESA activities investigate topics closely related to this work~\cite{esaact, esaact2}.
They examine Rust's applicability in space applications, including functional safety aspects and how to streamline the development of Rust in the presence of an existing C code base.

Ashmore et al. derived a set of evaluation criteria for software development in the air domain based on a software standard for aerospace systems~\cite{UKMOD}.
They formally evaluate Rust based on these criteria as a case study.
Their work focuses on the details of the programming language itself and not the ecosystem or practitioner-focused considerations.
They answer questions such as \textit{"How does the language prevent the introduction, or support the detection, of errors?"} by reviewing language features such as the Rust compiler's ownership model.

For our PowerPC Rust target case study, we draw inspiration from an experiment Ferrous Systems conducted in which they brought Rust support to bare metal SPARC v8 CPUs~\cite{FSSPARC}.

Willbold et al. performed a comprehensive analysis of the attack surface of satellite systems~\cite{odyssey}.
The authors provide a taxonomy for threats against embedded space systems and for the first time identify vulnerabilities in satellite firmware that allow remote takeover.

\section{Conclusion}
Our work demonstrates significant progress in integrating Rust into safety-critical space systems, addressing the urgent need for enhanced security and robustness in aerospace software development. 
We conduct an in-depth assessment of Rust's readiness for safety-critical applications, acting as a starting point for practitioners.
We fill the gap between C-for-Rust and Rust-for-C by introducing a practical methodology for gradually rewriting components in existing C-based systems in Rust.
The development of a new Rust compiler target for bare metal PowerPC CPUs further contributes to the applicability of Rust in an aerospace context.
Finally, the discovery and rectification of vulnerabilities in the Cubesat Space Protocol contribute to a safer status quo in open-source space software.

We hope that these efforts collectively help pave the way for a broader adoption of Rust in space systems, leading to a more secure, reliable, and safe aerospace software infrastructure.

\subsection{Availability}
Our contributions will be available on Github:
\url{https://github.com/pr0me/rust-for-critical-space-systems}.
We open-source our setup to fuzz libCSP with LibAFL, the partial Rust rewrite of libCSP and build instructions to link this partial rewrite back into the original C library as well as the \textit{rustc} target configuration for bare metal PowerPC.



\bibliographystyle{IEEEtranS}
\bibliography{input/mybib}

\end{document}